# Volumetric Particle Tracking Velocimetry (PTV) Uncertainty Quantification

**Sayantan Bhattacharya, Pavlos P. Vlachos**

Purdue University, Department of Mechanical Engineering, West Lafayette, USA

Abstract

We introduce the first comprehensive approach to determine the uncertainty in volumetric Particle Tracking Velocimetry (PTV) measurements. Volumetric PTV is a state-of-the-art non-invasive flow measurement technique, which measures the velocity field by recording successive snapshots of the tracer particle motion using a multi-camera set-up. The measurement chain involves reconstructing the three-dimensional particle positions by a triangulation process using the calibrated camera mapping functions. The non-linear combination of the elemental error sources during the iterative self-calibration correction and particle reconstruction steps increases the complexity of the task. Here, we first estimate the uncertainty in the particle image location, which we model as a combination of the particle position estimation uncertainty and the reprojection error uncertainty. The latter is obtained by a gaussian fit to the histogram of disparity estimates within a sub-volume. Next, we determine the uncertainty in the camera calibration coefficients. As a final step the previous two uncertainties are combined using an uncertainty propagation through the volumetric reconstruction process. The uncertainty in the velocity vector is directly obtained as a function of the reconstructed particle position uncertainty. The framework is tested with synthetic vortex ring images. The results show good agreement between the predicted and the expected RMS uncertainty values. The prediction is consistent for seeding densities tested in the range of 0.01 to 0.1 particles per pixel. Finally, the methodology is also successfully validated for an experimental test case of laminar pipe flow velocity profile measurement where the predicted uncertainty is within 17% of the RMS error value.

# Nomenclature

$x_w, y_w, z_w$: World coordinates or physical coordinates

$X^c, Y^c$: Camera image coordinates for camera c

$FX^c, FY^c$: $X$ and $Y$ calibration mapping function for camera c

$a_i$: camera mapping function coefficients

$e$: Error

$\sigma$: Standard uncertainty

$\Sigma$: Covariance matrix

$\vec{d}$: Disparity vector estimated from ensemble of reprojection error.

$u, v, w$: Velocity components in $x, y, z$ directions respectively.

$\Sigma_b$: Bias uncertainty

# 1 Introduction

Volumetric PTV (Maas, Gruen and Papantoniou, 1993; Baek and Lee, 1996; Ohmi and Li, 2000; Pereira *et al.*, 2006) is a fluid velocity measurement technique which resolves the three-dimensional (3D) flow structures by tracking the motion of tracer particles introduced in the flow. The tracer particle motion is recorded with multiple cameras to obtain projected particle images. Each camera is also linked to the physical space using a calibration mapping function(Soloff, Adrian and Liu, 1997). The particle images are then mapped back to the physical space using a triangulation process (Maas, Gruen and Papantoniou, 1993; Wieneke, 2008). Finally, a three-dimensional (3D) tracking of the reconstructed particles estimates the Lagrangian trajectories of the particles and subsequently resolves the volumetric velocity field. PTV easily lends itself to calculation of particle acceleration from the tracked trajectories. Also, unlike Tomographic Particle Image Velocimetry (Tomo-PIV) (Elsinga *et al.*, 2006), which involves spatial averaging over the interrogation window, 3D PTV has higher spatial resolution as it yields a vector for every tracked particle position. However, as the number of particles increases, identification of overlapping particles and its corresponding 3D reconstruction becomes challenging, which leads to a tradeoff between spatial resolution and reconstruction accuracy. Hence, the simple triangulation based 3D PTV method introduced in 1993 (Maas, Gruen and Papantoniou, 1993) had limited applications compared to Tomo-PIV for highly seeded flows. Improvements in terms of particle identification (Cardwell, Vlachos and Thole, 2011) and tracking algorithms (Cowen *et al.*, 1997; Takehara *et al.*, 2000; Riethmuller, 2001; Lei *et al.*, 2012; Fuchs, Hain and Kähler, 2016, 2017) have been proposed to minimize the error in the measurement.

Recent advancements in terms of reconstruction algorithms, such as Iterative Particle Reconstruction(IPR) (Wieneke, 2013) and Shake-the-box(STB) (Schanz, Gesemann and Schröder, 2016) have significantly improved the accuracy of 3D PTV. IPR uses an initial triangulation based reconstructed field to construct a projected image and then minimizes the intensity residuals in the image plane by shaking the particles in world coordinate location. This process achieves a better positional accuracy, reduced fraction of ghost particles and the reconstruction accuracy is comparable to intensity based Multiplicative Algebraic Reconstruction Technique (MART) (Elsinga *et al.*, 2006), for up to a seeding density of 0.05 particles per pixels (ppp). This concept has been further advanced in STB, which uses the temporal information, for a time-resolved measurement, to predict the particle location in the future frames and corrects the predicted position iteratively using IPR. Such measurements have successfully resolved flow structures for experiments with high particle concentrations (up to 0.125 ppp). With such capabilities, 3D PTV measurements have gained renewed attention and applicability in various experiments.

To analyze any experimental results with statistical significance, uncertainty quantification (UQ) is crucial, especially, where the measured data are used in a design process or to validate computational models (Angioletti, Nino and Ruocco, 2005; Ferreira, Van Bussel and Van Kuik, 2007; Ford *et al.*, 2008; van Ooij *et al.*, 2012; Brindise *et al.*, 2019). Given the increasing applicability and relevance of PTV/IPR/STB volumetric measurements, providing uncertainty estimation for an individual 3D PTV measurement is now of paramount importance.

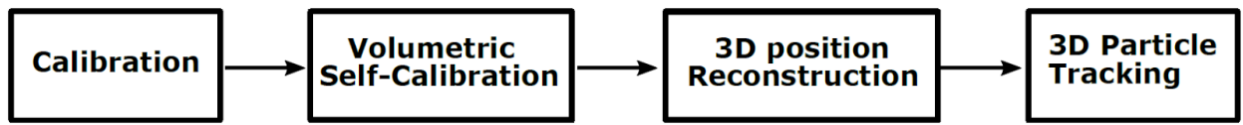

Figure 1: A volumetric PTV measurement chain showing the main steps in the process.

Uncertainty estimation in PIV measurements has received interest only recently and several methods have been proposed for planar PIV uncertainty quantification. Broadly such methods can be categorized into direct and indirect methods. Indirect methods rely on a calibration function, which maps an estimated measurement metric (e.g. correlation plane signal to noise ratio metrics (Charonko and Vlachos, 2013; Xue, Charonko and Vlachos, 2014, 2015) or estimates of the fundamental sources of error (Timmins *et al.*, 2012)) to the desired uncertainty values. Such a calibration is developed from a simulated image database and may not be sensitive to a specific error source for a given experiment. Direct methods, on the other hand, rely directly on the measured displacements and use the image plane "disparity" (Sciacchitano, Wieneke and Scarano, 2013; Wieneke, 2015) information or correlation-plane PDF (probability density function) of displacement information (Bhattacharya, Charonko and Vlachos, 2018) to estimate the a-posterior uncertainty values. Comparative assessments (Sciacchitano *et al.*, 2015; Boomsma *et al.*, 2016) have shown that the direct methods are more sensitive to the random error sources. However, indirect methods can be potentially used to predict any bias uncertainty. A direct uncertainty estimation for stereo-PIV measurement (Bhattacharya, Charonko and Vlachos, 2017) has also been proposed recently. A detailed review of such methods can be found in (Sciacchitano, 2019). Thus, although the foundations have been laid for planar and stereo-PIV uncertainty quantification, applicability of such methods to 3D measurements remains untested and these methods train strictly to cross-correlation based measurements. As a result, 3D reconstruction and tracking process for 3D PTV measurements is not covered under these methods and currently a-posterior uncertainty quantification methods for volumetric measurements (PTV/PIV) do not exist and new uncertainty model development is needed.

A flowchart for the different steps in a 3D PTV measurement chain is shown in Figure 1. The first step establishes a mapping function between the camera image coordinates $(X, Y)$ and the world coordinates $(x_w, y_w, z_w)$ in the physical space using a multi-camera calibration process. The calibration coefficients are then iteratively corrected using the mapping function and the recorded particle images to eliminate any misalignment between the assumed world coordinate system origin of the calibration plane and the actual origin location for the measurement volume. This process is called volumetric self-calibration(Wieneke, 2008) and is essential in minimizing the reconstruction error (due to existing offset or disparity between cameras) and improving the calibration accuracy. Using the modified calibration, for each particle in a given camera, the corresponding match in the second camera is searched along the epipolar line and the particle matches in all cameras are triangulated (Maas, Gruen and Papantoniou, 1993; Wieneke, 2008) to a 3D world position. This reconstruction process can be done in an iterative sense for an IPR type algorithm. However, for the particle pairing process in each camera view, the matching ambiguity increases for higher particle concentrations, which leads to erroneous reconstructions and is considered one of the main sources of error in the process. Finally, the reconstructed 3D particle positions are tracked to find the velocity vectors using "nearest neighbor" or other advanced algorithms (Fuchs, Hain and Kähler, 2017). The tracking and reconstruction can be done in conjunction for STB type evaluations. From calibration fitting error, particle position estimation error, the disparity vector estimation error to the error in finding the 3D positions and its pairing, the errors in each step of the process are inter-linked in a complex non-linear way and affect the overall error propagation. The iterative corrections and the governing non-linear functions lead to several interdependent error sources making the definition of a data reduction equation intractable and the development of an uncertainty quantification model non-trivial.

In the current framework, a model is developed to quantify the uncertainty in particle image position and the mapping function coefficient. These uncertainties are in turn combined with the uncertainty propagation through the reconstruction process. Finally, the uncertainty in the velocity vector is expressed directly as a combination of the position uncertainty in the matching pair of particles. The methodology is described in detail in the next section.

## 2 Methodology

The primary relation between the observed image coordinate $(X, Y)$ and the expected particle world coordinate $(x_w, y_w, z_w)$ in physical space is given by the individual camera mapping function $FX^c$ for each camera $c$, as given in equation (1).

$$X^c = FX^c(x_w, y_w, z_w, a_i) = a_1 + a_2 x_w + a_3 y + a_4 z + a_5 x_w^2 + a_6 x_w y_w + a_7 y_w^2$$

$$+ a_8 x_w z_w + a_9 y_w z_w + a_{10} z_w^2 + a_{11} x_w^3 + a_{12} x_w^2 y_w + a_{13} x_w y_w^2 \qquad (1)$$

$$+ a_{14} y_w^3 + a_{15} x_w^2 z_w + a_{16} x_w y_w z_w + a_{17} y_w^2 z_w + a_{18} x_w z_w^2 + a_{19} y_w z_w^2$$

Typically, a polynomial mapping function is used following Soloff et al. (Soloff, Adrian and Liu, 1997) to have higher accuracies in the presence of optical distortion effects. Once a mapping function is established and iteratively corrected using self-calibration process, the reconstruction process involves finding an inverse of the mapping function for the matching particle image coordinates in different projections. Hence an error propagation through the mapping function is the starting point of the uncertainty quantification and is described in the next subsection.

### 2.1 Error propagation through the mapping function

An error propagation for equation (1) can be written as follows:

$$e_{X^c} = \frac{\partial FX^c}{\partial x_w} e_{x_w} + \frac{\partial FX^c}{\partial y_w} e_{y_w} + \frac{\partial FX^c}{\partial z_w} e_{z_w} + \frac{\partial FX^c}{\partial a_i} e_{a_i} \qquad (2)$$

Equation (2) is obtained as a Taylor series expansion of equation (1), neglecting the higher order terms. Thus, the error in image coordinate $e_{X^c}$ can be related to the error in world coordinate positions $e_{x_w}$, $e_{y_w}$, $e_{z_w}$ and the error in calibration function coefficients $e_{a_i}$ through the mapping function gradients $\left(\frac{\partial FX^c}{\partial x_w}, \frac{\partial FX^c}{\partial y_w}, \frac{\partial FX^c}{\partial z_w}, \frac{\partial FX^c}{\partial a_i}\right)$. A similar propagation equation can be written for the error in $Y$ ($e_{Y^c}$) image coordinate for each camera mapping function. It is important to note that the quantities of interest are $e_{x_w}$, $e_{y_w}$, $e_{z_w}$ as we seek to estimate the unknown variance in the reconstructed world coordinate positions. Rearranging the unknown terms in the left-hand side and multiplying each side by its transpose yields the variance propagation equation as follows:

$$\left(\frac{\partial FX^c}{\partial x_w}e_{x_w} + \frac{\partial FX^c}{\partial y_w}e_{y_w} + \frac{\partial FX^c}{\partial z_w}e_{z_w}\right)\left(\frac{\partial FX^c}{\partial x_w}e_{x_w} + \frac{\partial FX^c}{\partial y_w}e_{y_w} + \frac{\partial FX^c}{\partial z_w}e_{z_w}\right)^T$$
$$= \left(e_{X^c} - \frac{\partial FX^c}{\partial a_i}e_{a_i}\right)\left(e_{X^c} - \frac{\partial FX^c}{\partial a_i}e_{a_i}\right)^T \quad (3)$$

The error in particle image position estimation ($e_{X^c}$) is a function of particle image fitting error and can be assumed to be independent of the error in calibration function coefficients ($e_{a_i}$). However, the calibration error can influence the error in projected particle image location or the projection error and thus any covariance between $e_{X^c}$ and $e_{a_i}$ is implicitly accounted in the projection error formulation, as discussed in section 2.2. With these considerations, a simplified version of equation (3) can be written as shown in equation (5).

$$\left[\frac{\partial FX^c}{\partial x_w} \quad \frac{\partial FX^c}{\partial y_w} \quad \frac{\partial FX^c}{\partial z_w}\right]\Sigma_{\vec{x}_w}\left[\frac{\partial FX^c}{\partial x_w} \quad \frac{\partial FX^c}{\partial y_w} \quad \frac{\partial FX^c}{\partial z_w}\right]^T = \sigma_{X^c}^2 + C_{\vec{a}}\Sigma_{\vec{a}}^c C_{\vec{a}}^T$$

**(4)**

Here, $\left[\frac{\partial FX^c}{\partial x_w} \quad \frac{\partial FX^c}{\partial y_w} \quad \frac{\partial FX^c}{\partial z_w}\right]$ is a row vector containing mapping function gradients for each camera $c$ with respect to $\vec{x_w} = \{x_w, y_w, z_w\}$ and $\Sigma_{\vec{x}_w}$ represents the unknown covariance matrix in world coordinates ($\Sigma_{\vec{x}_w} = \{e_{x_w}\ e_{y_w}\ e_{z_w}\}^T\{e_{x_w}\ e_{y_w}\ e_{z_w}\}$). The uncertainty in particle image position $X^c$ is denoted by $\sigma_{X^c}$. The term $C_{\vec{a}}\Sigma_{\vec{a}}^c C_{\vec{a}}^T$ evaluates to a single numerical value, which accounts for the contribution from the uncertainty in the calibration coefficients $\vec{a} = \{a_i\}_{1 \times 19}$, for the mapping function $FX^c$ of camera $c$. $C_{\vec{a}} = \left[\frac{\partial FX^c}{\partial a_i}\right]_{1 \times 19}$ represents the mapping function gradients with respect to the calibration coefficients $\vec{a}$ and the covariance in mapping function coefficients is denoted by

$\Sigma_{\vec{a}}^c = \{e_{a_i}\}\{e_{a_i}\}^T_{19 \times 19}$. For solving equation (4), it can be written as a stack of 8 rows of equations corresponding to $X$ and $Y$ mapping functions for each of, for example, a four-camera set-up. The combined equation for all cameras is given by equation (5) and is solved for each reconstructed particle individually.

$$C_{\vec{x}_w} \Sigma_{\vec{x}_w} C_{\vec{x}_w}^T = \Sigma_{\vec{X}} + \Sigma_{\vec{a}} \tag{5}$$

In equation (5), $C_{\vec{x}_w}$ is an 8x3 coefficient matrix containing mapping function gradients for the 8

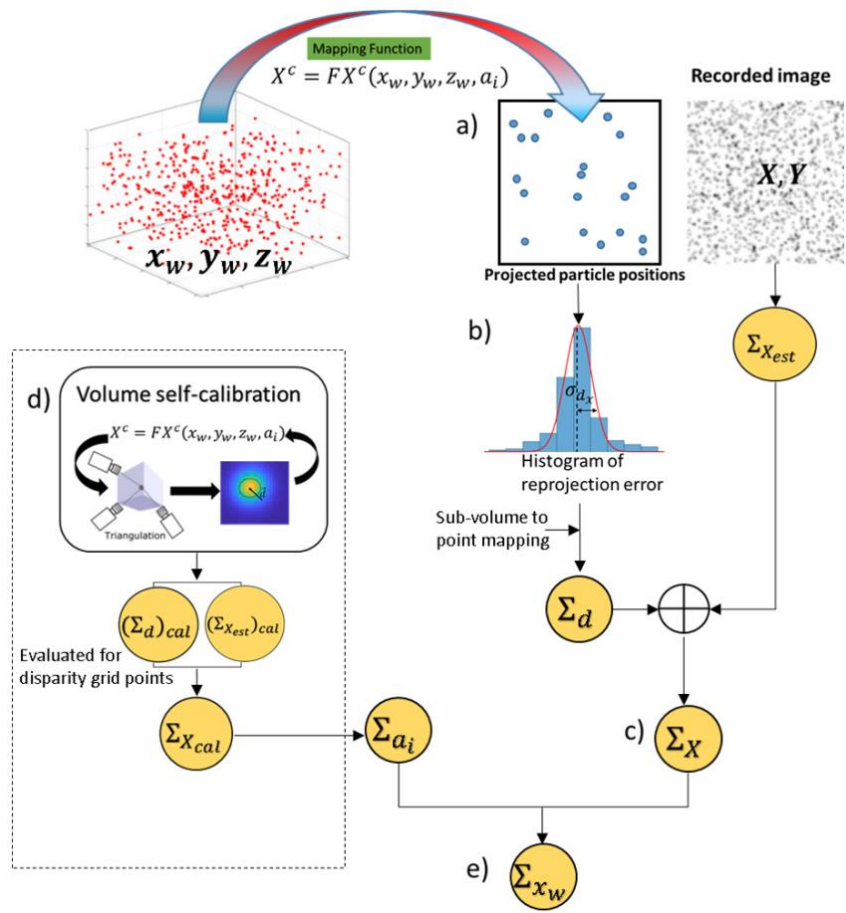

**Figure 2:** A schematic showing different steps (a – e) for estimating elemental uncertainties in particle image location $X$ and calibration coefficients $a_i$ and its propagation to the uncertainty in the world coordinate $x_w$.

mapping functions. The combined variance matrix in particle image position $\vec{X} = \{X^c, Y^c\}$ is denoted by $\Sigma_{\vec{X}}$ and contains $\sigma^2_{X^c}$ and $\sigma^2_{Y^c}$ as diagonal entries for each camera. The correlation in $e_{X^c}$ between different camera components is seen to be negligible and thus the off-diagonal terms of $\Sigma_{\vec{X}}$ are set to zero. Lastly, the evaluated values of $C_{\vec{a}} \Sigma^c_{\vec{a}} C^T_{\vec{a}}$ for each mapping function in equation (4) are put as the diagonal terms in the $\Sigma_{\vec{a}}$ matrix $((\Sigma_{\vec{a}})_{ii} = C_{\vec{a}} \Sigma^c_{\vec{a}} C^T_{\vec{a}})$, which represents the net calibration uncertainty contribution across all 4 cameras. Thus, equation (5) contains the unknown covariance matrix in world coordinates $\Sigma_{\vec{x}_w}$ as a function of $\Sigma_{\vec{X}}$ and $\Sigma_{\vec{a}}$. The following sections focus on estimating the $\Sigma_{\vec{X}}$ and $\Sigma_{\vec{a}}$ terms.

The overview of the uncertainty estimation and propagation process is depicted in Figure 2.

## 2.2 Estimating uncertainty in particle image location

For a-posteriori uncertainty quantification, we start from a reconstructed 3D particle positions obtained either from a triangulation or IPR reconstruction method. For a given 3D particle position, we want to find the corresponding projected particle image locations and its uncertainty for each camera. As shown in Figure 2a), the projected particle image positions are compared with the recorded image to find the error in particle image location. This can be expressed as a sum of the estimated projection error $(\vec{X}_{proj} - \vec{X}_{est})$ and the 2D particle fit position estimation error $(\vec{X}_{est} - \vec{X}_{true})$, for all $\vec{X} = \{X^c, Y^c\}$ and for each camera $c$, as shown in equation (6).

$$e_{\vec{X}} = \vec{X}_{proj} - \vec{X}_{true} = \vec{X}_{proj} - \vec{X}_{est} + \vec{X}_{est} - \vec{X}_{true} \qquad (6)$$

Thus, the variance in particle image location, $\Sigma_{\vec{X}}$, becomes a sum of the variance in the estimated projection error, denoted by $\Sigma_{\vec{a}}$, and variance of the error in particle image position estimation.

$$\Sigma_{\vec{X}} = e_{\vec{X}} e^T_{\vec{X}} = \Sigma_{\vec{a}} + \Sigma_{\vec{X}_{est}} \qquad (7)$$

As mentioned in section 2.1 equation (5), each of these variance matrices consider only the diagonal terms corresponding to $X$ and $Y$ mapping functions for each camera. In order to estimate $\Sigma_{\vec{a}}$ the reconstruction domain is divided into sub-volumes and the estimated projection error for a group of particles belonging to the same sub-volume are stacked up into a histogram (this relates to the concept of disparity$(\vec{d})$ defined by Wieneke (Wieneke, 2008)). The sub-volume size can be varied or particles from other frames can be included to have a larger statistical sample. It is observed that a histogram consisting of 50 or more particles in the sub-volume yields a statistically consistent estimate, irrespective of the number of sub-volumes considered. Such a histogram of disparity$(\vec{d})$ estimates is shown in Figure 2b), where the variance in the estimated $X$ projection error is denoted by $\sigma_{d_X}$. For a perfectly converged self-calibration, the mean disparity ($\bar{d}$) should be zero. Typically, the disparity histogram approaches a Gaussian distribution and for the robustness of variance estimation a Gaussian fit is applied on this histogram. The estimated standard deviation from the fitted curve is used to evaluate the variance of the disparity distribution. However, for a lower seeding density the disparity distribution is observed to deviate from a Gaussian distribution. Consequently, if the area under the fitted Gaussian curve is different by more than 5% compared to the histogram area evaluated using trapezoidal integration rule, the standard deviation of the distribution is used as the standard uncertainty. In this framework, this estimated variance is modeled as the desired $\Sigma_{\vec{d}}$ of equation (7). For the particles belonging to the same sub-volume, the same value of $\Sigma_{\vec{d}}$ is used.

Each particle image within $\pm 0.5$ pixels of the projected 3D particle location is fitted with a Gaussian shape and thus the uncertainty in the fitted position parameter for the least square fit process is considered as $\Sigma_{\vec{X}_{est}}$.

$$\Sigma_{\vec{X}_{est}} = (J^T J)^{-1} \sigma_{res}^2 I \qquad (8)$$

Equation 8 denotes an expression for the position estimation variance which is shown to be a function of the variance in the fit residual error ($\sigma_{res}^2$) and the Jacobian$(J)$ of the residual at the solution point (I denotes an identity matrix). This is consistent with the Cramer-Rao lower bound (CRLB) determination for 2D particle image centroid, as highlighted by (Rajendran, Bane and Vlachos, 2019). Hence, once $\Sigma_{\vec{d}}$ and $\Sigma_{\vec{X}_{est}}$ are estimated, the $\Sigma_{\vec{X}}$ is known (Figure 2c).

## 2.3 Estimating the uncertainty in mapping function coefficients

As seen from the flowchart in Figure 2, once the variance in particle image position$(\Sigma_{\vec{X}})$ is estimated through the progression of steps shown on the right side, the next workflow is focused on estimating the variance in the calibration coefficients ($\Sigma_{\vec{a}}$). The overall calibration uncertainty $\Sigma_{\vec{a}}$ is a combination of $\Sigma_{\vec{a}}^c$ for each camera $c$. The $\Sigma_{\vec{a}}^c$ estimation process (Figure 2d) can be performed in conjunction with the volumetric self-calibration process. In absence of self-calibration, the uncertainty in the coefficients $a_i$ is dictated by the uncertainty in calibration image dot fitting. However, the presence of disparity between estimated and projected points leads to a shift in the projected grid points $(X_{cal}, Y_{cal})$ in the image domain, and this correction leads to a new set of coefficients $(a_i)$ in the self-calibration process. Hence, the uncertainty in $X_{cal}, Y_{cal}$ positions, namely $\Sigma_{\overrightarrow{X_{cal}}}$, should directly affect the $\Sigma_{\vec{a}}^c$. If we consider the defined world coordinate positions $(x_{cal}, y_{cal}, z_{cal})$ where the disparity vectors are evaluated, then those grid points being chosen specific locations in space, will have no uncertainty in their location. Consequently, the unknowns $(e_{x_w}, e_{y_w}, e_{z_w})$ of equation (3) can be simplified to zero and the equation can be simplified to equation (9).

$$\Sigma_{\overrightarrow{X_{cal}}} = \left(\frac{\partial FX}{\partial a_i} e_a\right)\left(\frac{\partial FX}{\partial a} e_{a_i}\right)^T = C1_{\vec{a}} \Sigma_{\vec{a}}^c C1_{\vec{a}}^T \qquad (9)$$

In this equation, the $C1_{\vec{a}}$ represents the matrix of gradients of the mapping function with respect to the coefficients $a_i$, having number of rows corresponding to number of disparity grid points. The variance in the particle image position $\Sigma_{\overrightarrow{X_{cal}}}$ can be evaluated in a similar way as mentioned in section 2.22.1. Here, the $\Sigma_{\overrightarrow{X_{cal}}}$ can be evaluated for the initially triangulated particle positions and is used in equation (9) to solve for $\Sigma_{\vec{a}}^c$ as a least squares problem for all disparity grid points.

## 2.4 Uncertainty propagation in reconstructed positions

The uncertainty in the reconstructed world coordinate position is finally obtained by solving for the world coordinate location covariance matrix $\Sigma_{\vec{x}_w}$ from equation (5), as shown in Figure 2e). This equation is evaluated for each world coordinate position combining mapping functions in $X$ and $Y$ for all four cameras. The estimated covariance $\Sigma_{\vec{a}}^c$ term in section 2.3 is used to evaluate $C_{\vec{a}} \Sigma_{\vec{a}}^c C_{\vec{a}}^T$, where $C_{\vec{a}}$ represents $\frac{\partial FX^c}{\partial a_i}$ for each camera $c$, as mentioned in equation (4). The $\Sigma_{\vec{a}}$ term is then evaluated as a diagonal matrix as $(\Sigma_{\vec{a}})_{ii} = C_{\vec{a}} \Sigma_{\vec{a}}^c C_{\vec{a}}^T$. The $\Sigma_{\vec{X}}$ has already been calculated using equation (7). Hence, we solve for $\Sigma_{\vec{x}_w}$ by inverting the $C_{\vec{x}_w}$ matrix as shown in equation (10).

$$\Sigma_{\vec{x}_w} = B \left( \Sigma_{\vec{X}} + \Sigma_{\vec{a}} \right) B^{-1} \qquad (10)$$

Where, $B$ is given by $B = \left( C_{\vec{x}_w}^T C_{\vec{x}_w} \right)^{-1} C_{\vec{x}_w}^T$. It can be noted that for standard Gaussian particle images, the covariance between $X$ and $Y$ particle image position estimation can be assumed to be negligible. However, in presence of optical distortion, such a covariance can be estimated from the 2D least square fit of an elliptical Gaussian function on the mean particle image shape. Thus, the term $\left( \Sigma_{\vec{X}} + \Sigma_{\vec{a}} \right)$ is essentially an 8x8 diagonal matrix for 8 mapping function equations. From the covariance matrix $\Sigma_{\vec{x}_w}$, the standard uncertainty in reconstructed positions $(\sigma_{x_w}, \sigma_{y_w}, \sigma_{z_w})$ are obtained by taking the square root of the diagonal terms ($\sqrt{(\Sigma_{\vec{x}_w})_{ii}}$).

We also evaluate the bias uncertainty terms $\sigma_{x_b}, \sigma_{y_b}, \sigma_{z_b}$ based on the mean disparity value for each sub-volume. Ideally, for a converged self-calibration the mean disparity is negligible. However, due to measurement noise, any residual mean disparity ($\bar{d}$) can lead to a bias in the reconstructed position measurement. We estimate $\bar{d}$ from the disparity histogram and use that to estimate $\Sigma_{\vec{X}_b}$, the bias uncertainty in particle image position and $\Sigma_{\vec{a}_b}^c$, the bias uncertainty in $a_i$'s using the propagation equations (7) and (9). For $\Sigma_{\vec{X}_b}$, only $\Sigma_{\vec{a}_b}$ is considered in equation (7). The final bias uncertainty estimates for reconstructed $x, y, z$ positions are obtained using the propagation equation (10) by substituting the values of $\Sigma_{\vec{X}_b}$ and $\Sigma_{\vec{a}_b}$.

## 2.5 Uncertainty in estimated velocity field

The uncertainty in each tracked 3D velocity measurement is evaluated as a direct combination of the estimated 3D position uncertainties of each paired particle. Thus, if a particle in frame 1 $(\sigma_{x_{w1}}, \sigma_{y_{w1}}, \sigma_{z_{w1}})$ is paired with a particle in frame 2, then the uncertainty in the tracked displacement $\sigma_u$ is given by

$$\sigma_u^2 = \sigma_{x_b}^2 + \sigma_{x_{w1}}^2 + \sigma_{x_{w2}}^2 - \rho_{x_{w1}x_{w2}}\sigma_{x_{w1}}\sigma_{x_{w2}} \qquad (11)$$

In equation (11), $\sigma_{x_b}$ is the bias uncertainty term as evaluated in section 2.4. The bias uncertainty depends on the mean disparity and the mapping function coefficients and is not expected to change from frame to frame. Hence it is accounted for only once in the tracking uncertainty estimation. It is also observed that the true position error in the estimated 3D particle position for a paired particle in frame 1 and frame 2 has a strong correlation. Thus, the covariance term $\rho_{x_{w1}x_{w2}}\sigma_{x_{w1}}\sigma_{x_{w2}}$ in equation (11) is significant. The correlation coefficient $\rho_{x_{w1}x_{w2}}$ varies from about 0.5 to 0.8, depending on the flow field and calibration and is estimated as an average of the correlation of the individual camera disparity error between paired particles. The $\rho_{x_{w1}x_{w2}}$ term can be computed for each pair of frames and also for a statistically significant number of particles within the same sub-volume. However, if the spatio-temporal variations of $\rho_{x_{w1}x_{w2}}$ is within 5% of the mean value, then an average coefficient may be used to calculate the covariance term. The disparity error correlation is expected to have a similar magnitude compared to the true position error correlation between frames and is verified to be the case for synthetic test cases with true error quantification. The uncertainty in $v$ and $w$ components ($\sigma_v, \sigma_w$) can be obtained in a similar way following equation (11). It is to be noted, that the uncertainty due to false matching in presence of ghost particles may need further analysis. However, for a valid measurement we expect equation (11) to account for the uncertainty in the tracked velocity measurement.

## 3 Results

The proposed framework to estimate the uncertainty in the reconstructed particle positions is tested using synthetic vortex ring images. The particle field was generated and advected using incompressible axisymmetric vortex ring equations mentioned in (Wu, Ma and Zhou, 2006). The camera calibration and particle images (256x256 pixels) were generated using in-house code. The camera angles were selected as 35° and were positioned in a plus (+) configuration. The volume of interest was set to 42mmx42mmx24mm and the seeding density was varied from 0.01ppp to 0.1ppp. The processing was also done using in-house calibration and IPR code for 100 image pairs. A polynomial model was used for the camera calibration and the initial estimate of the calibration was modified by 3 iterations of volumetric self-calibration to eliminate any mean disparity. An allowable triangulation error of 1 pixel was used for initial triangulation with particle identification using dynamic particle segmentation method (Cardwell, Vlachos and Thole, 2011) to better resolve overlapping particle images. The particle image positions were estimated using least square Gaussian fit. The optical transfer function (OTF) (Schanz *et al.*, 2013) was calculated and used in IPR iterations. The number of inner loop and outer loop iterations for each frame was set to 4 with particle "shaking" of $\pm 0.1$ voxels. The 3D particle tracking was done using "nearest neighbor" algorithm. The uncertainty for each measurement was computed using the set of equations described in section 2.

### 3.1 Comparing error and uncertainty histogram for reconstructed particle positions

First, the uncertainty in reconstructed particle positions are analyzed. The reconstructed particle positions are compared with the true particle positions in space and if a particle is found within 1 voxel radius of the true particle, then it is considered as a valid reconstruction. The error in reconstructed $x_w$ position is denoted by $e_{x_w}$ and defined as:

$$e_{x_w} = x_w^{estimated} - x_w^{true} \tag{12}$$

Similarly, $e_{y_w}$ and $e_{z_w}$ are defined. Figure 3 shows the histogram of error and uncertainty distributions $x_w, y_w$ and $z_w$ coordinates. Figure 3a and Figure 3b shows the distributions for the reconstructed particle positions obtained using triangulation and IPR methods respectively, for a particle concentration of 0.05ppp. The x-axis is divided into 60 equally spaced bins and the y-axis denotes the number of measurements falling within each bin as a fraction of total number of points. The root mean squared (RMS) error is defined as:

$$RMS\ error = \sqrt{\frac{1}{N}\sum_{i=1}^{N} e_{i_w}^2} \tag{13}$$

The error distribution for the triangulated particle positions is wider with RMS error of about 0.17, 0.18 and 0.27 pixels in $x_w, y_w$ and $z_w$ positions compared to RMS error of 0.15, 0.15 and 0.22 pixels for the IPR case. The predicted uncertainty distributions have significantly less spread and have a tight distribution around the RMS error. For a successful prediction, it is expected that the RMS value of the error distribution should match the RMS value of the estimated uncertainty distribution (Sciacchitano *et al.*, 2015). The RMS value for each distribution is indicated by the dashed vertical line. For Figure 3a, the RMS uncertainty values underpredict the RMS error by 0.03 pixels in $x_w$ and $y_w$ and by 0.06 pixels in $z_w$. For IPR case in Figure 3b, the predicted uncertainties are within 0.02 pixels of the RMS error values. Overall, the predicted uncertainties are in close agreement with the expected value, indicating a successful prediction for position reconstruction uncertainty.

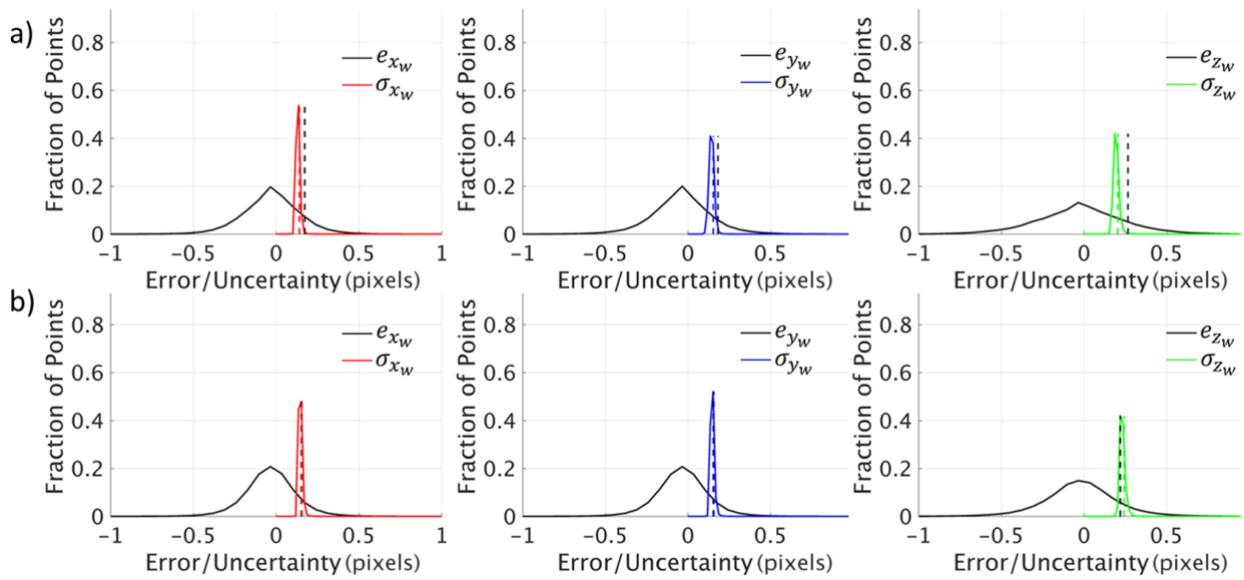

**Figure 3: Histogram of error ($e$) and uncertainty ($\sigma$) distributions for reconstructed particle positions ($x_w, y_w, z_w$) for the synthetic vortex ring case with 0.05ppp particle concentration for a) triangulation and b) IPR reconstructions. The vertical lines indicate the RMS value for each distribution.**

## 3.2 Reconstructed position uncertainty for varying particle concentration

The increase in particle concentration leads to a higher percentage of overlapping particles which increases the error in particle identification, and subsequently in 3D particle reconstruction. To test the sensitivity of the uncertainty predictions in such scenarios, the seeding density is varied from 0.01ppp to 0.1ppp and the RMS error and uncertainty values are compared in each case, as shown in Figure 4a and Figure 4b. The results show a high sensitivity of the predicted uncertainty to the trend of the RMS error for both triangulation and IPR methods. The reconstructed position RMS error predicted by IPR is lesser than the triangulation error for lower seeding densities, whereas, for 0.1ppp the IPR error is higher, which may be related to the specifics of the in-house IPR implementation. However, the objective is to predict the correct RMS error level given the different reconstructed positions using different methodologies. For triangulation the RMS uncertainty follows the RMS error trend consistently, but underpredicts the magnitude by about 0.04 pixels

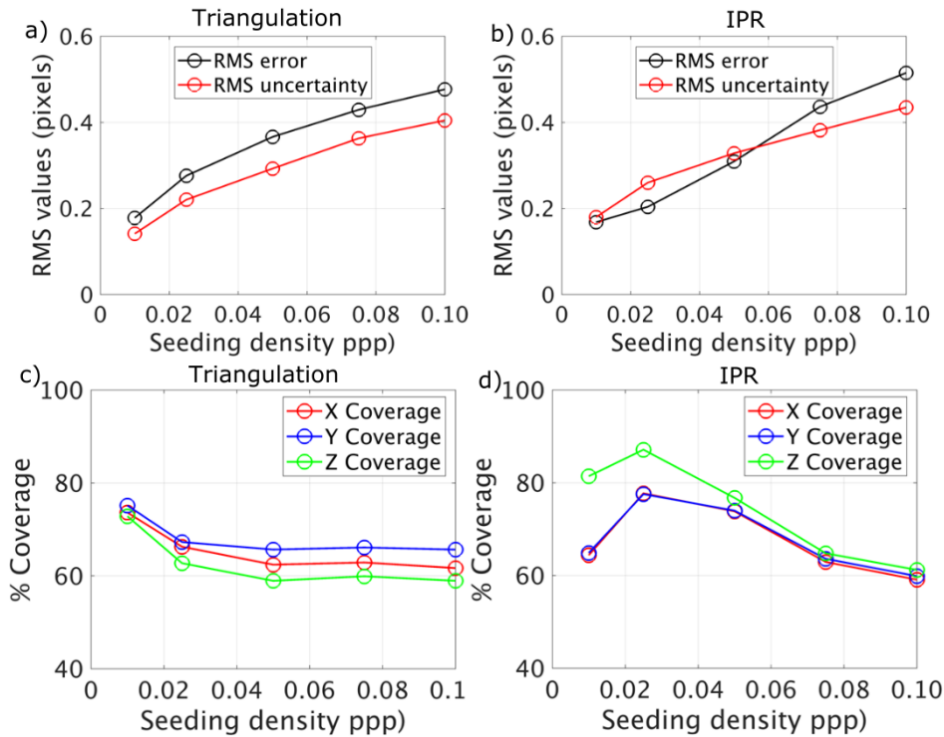

**Figure 4: Comparison of triangulation and IPR reconstructed position error and uncertainty as a function of seeding density for the synthetic vortex ring case. Plot a) compares the RMS error and RMS of predicted uncertainties for seeding densities in the range of 0.01ppp to 0.1ppp and plot b) compares the coverage in each case.**

(23%) at 0.01ppp and by 0.07 pixels (20%) at 0.1ppp. For the IPR case, the predicted uncertainty matches the expected uncertainty value closely at 0.01ppp and 0.05ppp with a deviation of about 0.01 pixels (10%), but underpredicts the uncertainty by 0.08 pixels (30%) at 0.1ppp. Overall the increasing trend agreement, between the predicted and the expected uncertainty validates the current framework for prediction of uncertainty for a wide range of particle concentrations and using both reconstruction methods.

**Table 1: Comparison of RMS error and RMS uncertainty values for the triangulation and IPR based reconstructed particle positions for a range of seeding densities.**

| Particle Concentration (ppp) | RMS $e_{x_w}$ (voxels) | RMS $\sigma_{x_w}$ (voxels) | RMS $e_{y_w}$ (voxels) | RMS $\sigma_{y_w}$ (voxels) | RMS $e_{z_w}$ (voxels) | RMS $\sigma_{z_w}$ (voxels) |
|---|---|---|---|---|---|---|
| Triangulation Reconstruction | | | | | | |
| 0.010 | 0.08 | 0.07 | 0.09 | 0.07 | 0.13 | 0.10 |
| 0.025 | 0.13 | 0.11 | 0.14 | 0.11 | 0.20 | 0.16 |
| 0.050 | 0.17 | 0.14 | 0.18 | 0.15 | 0.27 | 0.21 |
| 0.075 | 0.21 | 0.17 | 0.22 | 0.19 | 0.31 | 0.25 |
| 0.100 | 0.23 | 0.19 | 0.24 | 0.22 | 0.34 | 0.28 |
| IPR Reconstruction | | | | | | |
| 0.010 | 0.09 | 0.08 | 0.09 | 0.09 | 0.11 | 0.13 |
| 0.025 | 0.11 | 0.12 | 0.11 | 0.12 | 0.14 | 0.19 |
| 0.050 | 0.15 | 0.16 | 0.15 | 0.16 | 0.22 | 0.24 |
| 0.075 | 0.22 | 0.18 | 0.22 | 0.18 | 0.31 | 0.28 |
| 0.100 | 0.26 | 0.21 | 0.26 | 0.21 | 0.36 | 0.32 |

For a more specific comparison across seeding densities, the values of RMS errors and uncertainties in $x_w, y_w$ and $z_w$ positions for both methods have been presented in Table 1. The maximum underprediction of about 0.06 pixels occurs at 0.1ppp case for both methods. The best agreement is obtained for the IPR case for up to 0.05ppp and for the triangulation case upto 0.025ppp. It is to be noted that the IPR reconstruction error is higher than expected, which may be related to a lower convergence rate and in turn depends on the specifics of the implementation, however, given a reconstructed field the current method reasonably predicts the standard uncertainty in 3D particle based reconstruction.

To compare the global prediction of uncertainty level for all particles the estimated coverage is plotted in Figure 4c and Figure 4d. The coverage is defined as the percentage of measurement errors falling within the uncertainty bound ($\pm\sigma$). For an ideal Gaussian error distribution, the standard uncertainty coverage is 68.3%. In Figure 4c, the coverage for all cases lies within 60% to 68%, except for 0.01 ppp for which case the coverage is about 74% for triangulation. The deviation for lower seeding density case may be related to the non-Gaussian nature of the error distributions at such particle concentrations. For IPR the coverage varies from 60% to 87%, with maximum overprediction for the 0.025ppp case, as shown in Figure 4d. Thus, the uncertainty coverage metric is mostly in the range of 60% to 73% in the present analysis and agrees well with the ideal expected coverage of 68.3%.

**Table 2: Comparison of RMS error and RMS uncertainty values for the particle tracking displacement estimates using triangulation and IPR based reconstructed particle positions for a range of seeding densities.**

| Particle Concentration (ppp) | RMS $e_u$ (voxels /frame) | RMS $\sigma_u$ (voxels /frame) | RMS $e_v$ (voxels /frame) | RMS $\sigma_v$ (voxels /frame) | RMS $e_w$ (voxels /frame) | RMS $\sigma_w$ (voxels /frame) |
|---|---|---|---|---|---|---|
| Triangulation Reconstruction | | | | | | |
| 0.010 | 0.06 | 0.05 | 0.07 | 0.05 | 0.10 | 0.07 |
| 0.025 | 0.09 | 0.08 | 0.09 | 0.08 | 0.14 | 0.11 |
| 0.050 | 0.11 | 0.11 | 0.12 | 0.12 | 0.18 | 0.16 |
| 0.075 | 0.13 | 0.15 | 0.14 | 0.16 | 0.20 | 0.21 |
| 0.100 | 0.14 | 0.17 | 0.15 | 0.19 | 0.22 | 0.25 |
| IPR Reconstruction | | | | | | |
| 0.010 | 0.12 | 0.08 | 0.12 | 0.09 | 0.13 | 0.13 |
| 0.025 | 0.12 | 0.11 | 0.12 | 0.11 | 0.14 | 0.17 |
| 0.050 | 0.16 | 0.14 | 0.16 | 0.14 | 0.22 | 0.22 |
| 0.075 | 0.22 | 0.18 | 0.22 | 0.18 | 0.30 | 0.28 |
| 0.100 | 0.25 | 0.21 | 0.25 | 0.22 | 0.34 | 0.33 |

### 3.3 Uncertainty prediction for tracked velocity vectors

As a final step, the uncertainty prediction in the tracked velocity field is assessed. The reconstructed 3D particle positions are tracked for a pair of frames for 100 pairs using nearest-neighbor tracking. The true particle positions in 1 voxel vicinity of the reconstructed particle positions is found for the first frame and the corresponding true displacement is subtracted from the estimated displacement to compute the error $(e)$ in $u$, $v$ and $w$ velocity components. A measurement is considered valid if the computed error magnitude is within 1 voxel. The uncertainty $(\sigma_u, \sigma_v, \sigma_w)$ in the velocity components are computed using equation (11).

The RMS uncertainty values mentioned in Table 2 are in close agreement with the RMS error values with a maximum deviation of 0.04 pixels across all cases. The RMS error increases with the particle concentration due to higher probability of erroneous matches resulting from ghost particle reconstruction. The predicted uncertainty increases proportionally with RMS error, for both reconstruction methods, as observed in Table 2.

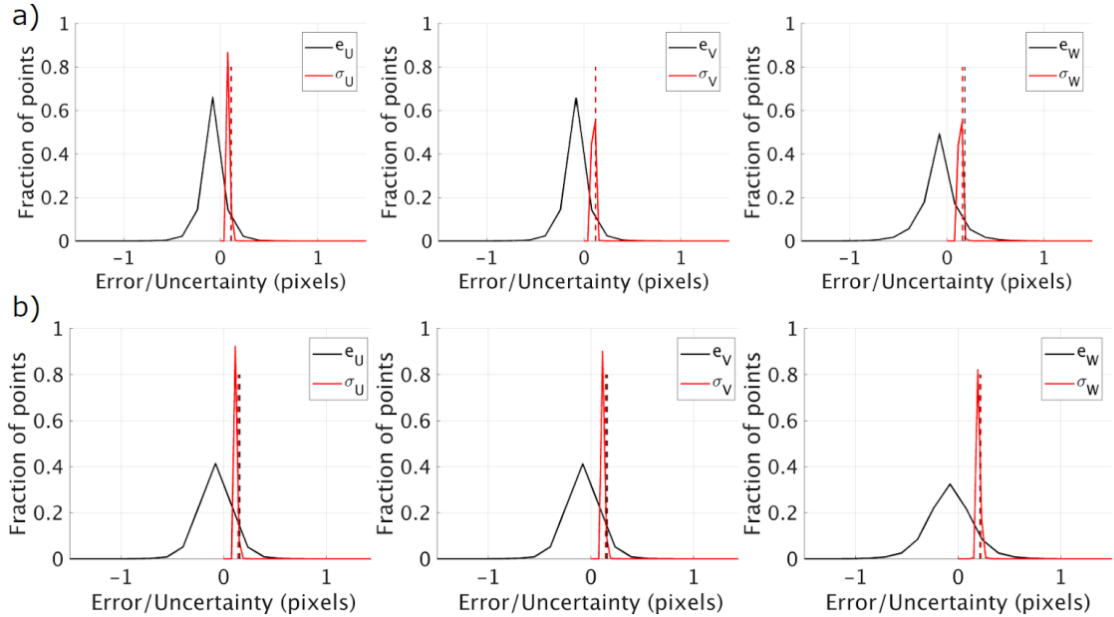

**Figure 5: Error and uncertainty histogram comparison for tracked velocity vectors in the synthetic vortex ring case with seeding density of 0.05ppp for a)triangulation based reconstruction and for b)IPR based reconstruction.**

The histogram of velocity error and uncertainty distribution is compared in Figure 5a for the triangulation case and Figure 5b for the IPR case, for 0.05ppp seeding density. The error distribution is sharper for the triangulation case. It is noticed that the $w$ component has higher error compared to $u$ and $v$ components. For all cases, the uncertainty distributions have a very narrow spread and predicts the RMS error magnitude perfectly. Further analysis for higher seeding densities with STB processing is required to validate the displacement uncertainty model proposed by equation (11), however, these results show reasonable agreements between predicted and expected uncertainty values for the estimated velocity components..

## 3.4 Experimental Validation: Uncertainty prediction for laminar pipe flow

The current framework is also validated for a canonical laminar pipe flow experiment for a Reynolds number of 630. The schematic of the experimental set up is shown in Figure 6. The flow loop consisted of a gear pump driving a steady flow rate of 0.17 L/min through a circular FEP tube of 0.25 inches diameter. The working fluid inside the pipe was chosen as distilled water-urea (90:10) solution with a density of 1015 kg/m3 and dynamic viscosity of 0.915 mPas. The tube was fully immersed in an acrylic tank filled with water-glycerol solution such that it is refractive index matched. The volumetric PTV measurement was performed using four Phantom Miro M340 cameras with three cameras at the same horizontal plane and one camera angled in the vertical plane, as shown in the sideview of Figure 6. The flow rate in the upstream and downstream of the pipe was measured using an ultrasonic flowmeter and the average flow rate was used to determine the true velocity profile. The measurement volume was 9x6.5x6.5 mm3 and was illuminated by a continuum Terra-PIV laser with appropriate optical setup. The time-resolved measurements were taken at 6 kHz, and the image size was 640x624 pixels with an average magnification of 17.8 microns/pixel. 24-micron fluorescent particles were used with a particle Stokes number St= 0.0005. The particle images were processed using in-house camera calibration, particle reconstruction and tracking code. A polynomial mapping function (Soloff, Adrian and Liu, 1997) was used to establish a relation between image coordinates and physical coordinates. Three iterations of volumetric self-calibration (Wieneke, 2008) were done to eliminate any disparity between the measurement volume and calibration target location or alignment. Both triangulation and IPR was used to reconstruct the particle positions in physical coordinate system and subsequently the 3D particle locations were tracked using a "nearest-neighbor" pairwise tracking algorithm. 500 pairs of images were processed with a particle concentration of 0.005ppp.

The reconstructed particle positions across all images are summed up in the cross-sectional view of the tube and a least square circular fit is performed to fit a circle with size closest to the diameter of the tube. The fitted boundary is used to divide the cross-sectional area of the tube in 20x20 bins and all measurements in streamwise direction as well as across 500 frames are averaged per bin to obtain the mean velocity profile shown in Figure 7a. The mean velocity profile along the middle y-plane is compared with the true solution in Figure 7b. The expected true velocity profile $U_{true}$ for the measured flow rate is shown by the blue solid line. The flow meter has a 10% uncertainty and its corresponding standard uncertainty ($\pm\sigma$) is shown by the blue shaded region. The mean velocity profile obtained from particle tracks (for the triangulation case) is shown by the black solid line and the standard deviation of the velocity measurements in each bin is shown by the shaded grey region. The peak measured velocity reaches 94% of the true maximum velocity. The standard deviation of the measured velocity is observed to increase in the depth direction moving away from the camera. Overall, the mean velocity profile agreed with the expected parabolic profile of a laminar pipe flow.

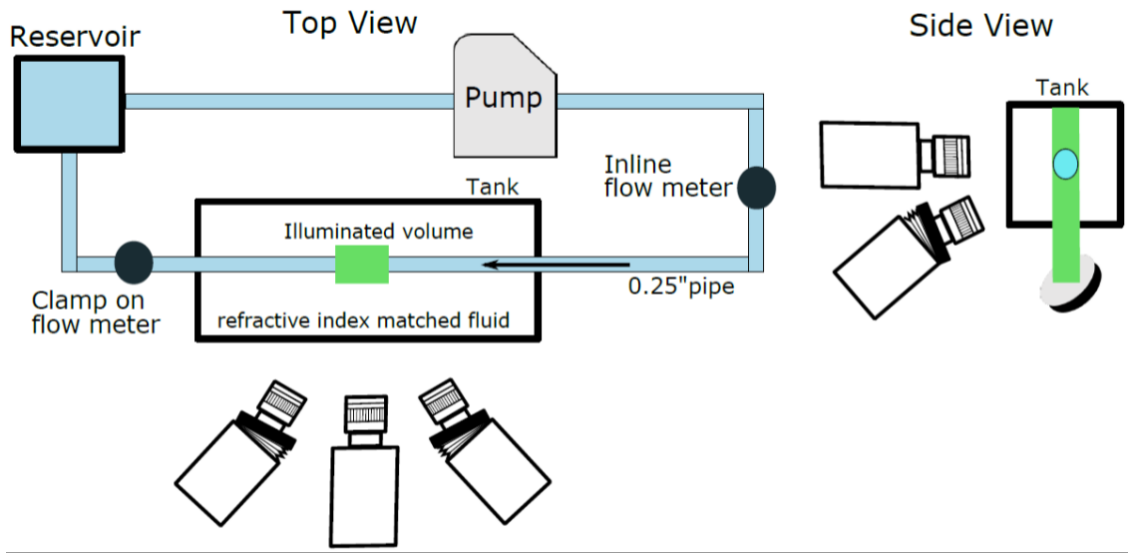

**Figure 6: Schematic of laminar pipe flow set up showing the flow loop and camera arrangement.**

The measured streamwise component of velocity ($U$) is compared with the true expected velocity ($U_{true}$) and the distribution of velocity tracking error $e_U$ and the estimated corresponding uncertainty $\sigma_U$ is shown in Figure 7c and Figure 7d for the triangulation and IPR reconstruction cases respectively. In both cases the error distribution is skewed with a higher bias error for the triangulation case of about 0.1 pixels/frame. The predicted uncertainty values are distributed closely about the RMS error value. The RMS error and RMS uncertainty values for Figure 7c are 0.17 pixels/frame and 0.14 pixels/frame and for Figure 7d are 0.23 pixels/frame and 0.19 pixels/frame respectively. Thus, the predicted uncertainty using the current framework shows 0.04 pixels underprediction and reasonably predicts the appropriate measurement uncertainty level.

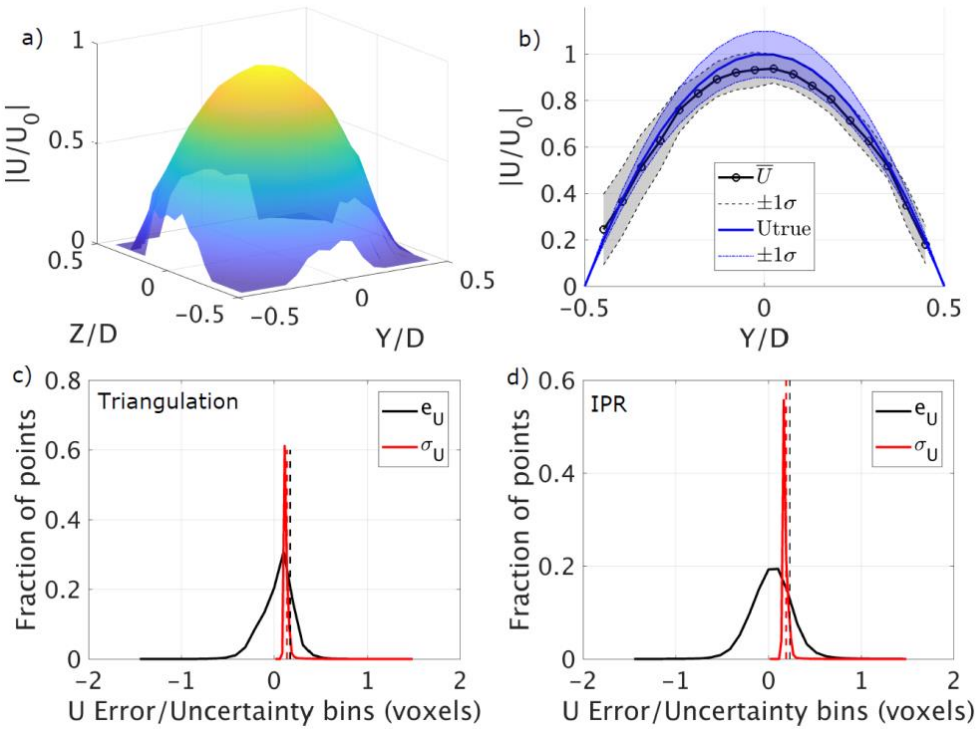

**Figure 7: The mean streamwise velocity profile for a 3D PTV measurement of a laminar pipe flow is shown in a). The velocity profile is compared with the true solution in b). The error and estimated uncertainty histogram are shown for triangulation-based reconstruction in c) and for IPR based reconstruction in d).**

## 4   Conclusion

We presented a comprehensive framework to predict the uncertainty in the reconstructed 3D particle positions in a volumetric PTV measurement and subsequently propagate the uncertainty in the tracked velocity estimates. The variance estimated from the histogram of the projection error provides the uncertainty bound on the particle image position and contributes to the uncertainty in the mapping function coefficients. The uncertainty on the reconstructed 3D position is obtained as a combination of the particle image position uncertainty and the mapping function coefficient uncertainty. The bias uncertainty on the reconstructed particle positions due to the residual mean disparity is also considered. For the tracked velocity uncertainty, the uncertainty in the reconstructed particle positions is directly combined for each matching particle pair. The covariance between particle position error for paired particles in frame 1 and frame 2 is estimated using the correlation coefficient of the disparity error values for corresponding particles. Analysis with the synthetic vortex ring images showed good agreement between the RMS of the predicted uncertainties in $x_w, y_w, z_w$ positions and the RMS error. The estimated uncertainty in the displacement field was within 0.04 voxels/frame of the RMS error for both the vortex ring case and the experimental pipe flow case. Overall, the predicted uncertainties are sharply distributed close to the RMS error values and showed strong sensitivity to the variation in RMS error, across a range of seeding densities.

The proposed methodology is applicable, in general, for any given set of 3D reconstructed particle positions, even when they are obtained using advanced tracking methods like STB. However, for STB, the uncertainty in particle trajectory fitting should also be quantified. The current methodology assumes negligible variance in laser pulse separation and thus ignores any temporal uncertainty in the particle tracking. The method also assumes that any covariance in particle image position and calibration coefficient is implicitly taken into account by the uncertainty in the projection error. Another key assumption in this process is the independence between $X$ and $Y$ particle image position estimation errors. These limitations can be further explored and the covariance terms can be quantified in future work. The distinction of uncertainty levels for true and false reconstructions should also be further analyzed to explore uncertainty predictions for ghost particle reconstructions. In conclusion, the proposed framework demonstrates accurate uncertainty predictions for both the vortex ring and the pipe flow test cases. These results establish the current methodology as the first successful predictor for uncertainty in a 3D PTV measurement.

# 5 Acknowledgement

This work is supported by the National Science Foundation's MRI program grant with award number 1725929.